\documentclass[12pt,preprint]{aastex}

\usepackage{color}

\newcommand{\ptsec}{\mbox{\ensuremath{.\!\!^{\prime\prime}}}}

\slugcomment{Submitted to ApJ}

\shorttitle{H$_2$ Emission from GG Tau A}
\shortauthors{Beck et al.}

\begin{document}


\title{Circumbinary Gas Accretion onto a Central Binary:  Infrared Molecular Hydrogen Emission from GG Tau A}


\author{Tracy L. Beck\altaffilmark{1}, Jeffrey. S. Bary\altaffilmark{2},  Anne Dutrey\altaffilmark{3}, Vincent Pi\'etu\altaffilmark{4}, Ste\'phane Guilloteau\altaffilmark{3}, S. H. Lubow\altaffilmark{1},  M. Simon\altaffilmark{5}, }

\altaffiltext{1} {The Space Telescope Science Institute, 3700 San Martin Dr. Baltimore, MD 21218}
\altaffiltext{2} {Colgate University, Department of Physics \& Astronomy, 13 Oak Drive, Hamilton, NY  13346}
\altaffiltext{3} {Universit\'e de Bordeaux, Observatoire Aquitain des Sciences de l$'$Univers (OASU), 2 rue de l$'$Observatoire, BP89, F-33271 Floirac Cedex, France }

\altaffiltext{4} {IRAM, 300 Rue de la Piscine, 38406 Saint Martin d$'$H\'eres, France}
\altaffiltext{5} {Stony Brook University, Department of Physics and Astronomy, Stony Brook, NY 11794}

\email{tbeck@stsci.edu, jbary@colgate.edu, Anne.Dutrey@obs.u-bordeaux1.fr,  pietu@iram.fr, Stephane.Guilloteau@obs.u-bordeaux1.fr, lubow@stsci.edu, michal.simon@sunysb.edu}




\begin{abstract}

We present high spatial resolution maps of ro-vibrational molecular hydrogen emission from the environment of the GG~Tau~A binary component in the GG~Tau quadruple system.  The H$_2$ {\it v}~=~1-0~S(1) emission is spatially resolved and encompasses the inner binary, with emission detected at locations that should be dynamically cleared on several hundred-year timescales.  Extensions of H$_2$ gas emission are seen to $\sim$100~AU distances from the central stars.  The {\it v}~=~2-1~S(1) emission at 2.24~$\mu$m is also detected at $\sim$30~AU from the central stars, with a line ratio of 0.05$\pm$0.01 with respect to the  {\it v}~=~1-0~S(1) emission.  Assuming gas in LTE, this ratio corresponds to an emission environment at $\sim$1700~K .  We estimate that this temperature is too high for quiescent gas heated by X-ray or UV emission from the central stars.  Surprisingly, we find that the brightest region of H$_2$ emission arises from a spatial location that is exactly coincident with a recently revealed dust ``streamer" which seems to be transferring material from the outer circumbinary ring around GG~Tau~A into the inner region.  As a result, we identify a new excitation mechanism for ro-vibrational H$_2$ stimulation in the environment of young stars.  The H$_2$ in the GG~Tau~A system appears to be stimulated by mass accretion {\it infall} as material in the circumbinary ring accretes onto the system to replenish the inner circumstellar disks.  We postulate that H$_2$ stimulated by accretion infall could be present in other systems, particularly binaries and``transition disk" systems which have dust cleared gaps in their circumstellar environments.

\end{abstract}


\keywords{physical data and processes: accretion, accretion disks --- stars: pre-main sequence --- stars: winds, outflows, accretion --- stars: formation ---  planetary systems: protoplanetary disks --- stars: individual (GG Tau A)}



\section{Introduction}

Planet searches and corresponding imaging follow-up suggest that as many as 25\% of  extrasolar planets reside in binary or higher order multiple stellar systems \citep[e.g.][]{pati2002,egge2004,mugr2007,take2008,mugr2009}.    This multiplicity fraction is likely a lower limit, since most searches have preferentially selected the potential hosts to be single star systems.  Such a high multiplicity fraction for exoplanet host systems indicates that despite the increased dynamical complexity of young multi-stellar systems, which lead to the truncation of individual circumstellar disks and the rapid clearing ($\le$1000~yr) of the inner regions of circumbinary(system) disks, reservoirs of material frequently survive long enough for planets to form.  Stars commonly form in binary systems, so the fact that planets exist in these systems adds tremendously to the diversity and prevalence of planets in the Galaxy \citep{duqu1991,ghez1993,reip1993,lein1993,simo1995,duch2007}.

Gas flows through dust-cleared gaps in the disks of binary star systems can also provide important clues about the processes that could occur in the embedded planetary systems that open gaps. Planetary systems can be considered to be binaries of extreme mass ratios.  Most of the mass gain of gas giant planets may occur by means of mass flow through the gap surrounding the orbit of the planet  \citep{lubo1999,lubo2006,ayli2010}  The gas flow through gaps in young planetary systems and onto circum-planetary disks plays a crucial role in the formation of the young planets.  The physics of this process can be better understood with direct observational constraints placed on the similar processes that occur in young binary star systems.

In order to understand the formation of planets in young binary and multiple systems, we first must understand the ways in which the stellar cores interact with the gas and dust in their circumstellar environment.  Clues to the dynamics of these interactions and the structures in the circumstellar/system environment are revealed by spectral energy distribution measurements, high-spatial resolution observations, and hydrodynamic simulations  \citep{jens1997,arty1994,dutr1994,arty1996,gunt2002,pasc2008}.  For instance, simulations showing how young binary systems quickly clear gaps in the inner regions of circumbinary disks also describe how material will fall through these gaps in ``high-velocity streams,'' eventually crashing into the individual circumstellar disks \citep{arty1996,gunt2002}.  High-spatial resolution millimeter and optical/near-infrared observations of nearby young binary systems like GG~Tau~A clearly show that optically-thin regions exist between the circumbinary disk and the stellar components \citep{dutr1994,guil1999,rodd1996,mcca2002,kris2002,kris2005,piet2011}.   However, these observations also strongly hint at the presence of streamers of material falling through the optically-thin, dynamically unstable region \citep{piet2011}.  More recently, a near-infrared high-resolution image of scattered light from SR~24, a binary system in nearby the Ophiuchus star forming region, clearly indicates the presence of circumstellar disks orbiting both the primary and secondary and a spiral arm-like stream of material extending some 1600~AU from central sources toward a circumbinary disk \citep{maya2010}.

The replenishment of the tidally-truncated, inner circum{\it stellar} disks with material from the extended circum{\it binary(system)} disks is likely to aid the formation of planets in many binary and multiple systems.   Moreover, the recent discovery of Kepler-16b, an exoplanet that lies within the same orbital plane as its central binary host, demonstrates the potential for planets to form within circumbinary disks \citep{doyl2011}.  High-spatial resolution images capable of resolving emission from structures within a circumbinary region and near the inner circumstellar disks will be essential for studying the planet-building process in such systems.   Constraining existing hydrodynamical models, which form our basis for understanding the evolution of young multiple systems and their host planets, would be a meaningful step in unraveling the complexity of star and planet formation in the binary environment.

As an exciting step toward this goal, we present the first spatially resolved image of molecular hydrogen gas located within the optically-thin, unstable region of the GG~Tau~A system.  This image provides strong support for the existence of suspected accretion ``streamers'' in this system.  The material falling onto the inner ambient gas appears to create a shock front located in the dynamically unstable circumstellar environment of the stars.   We compare the ratios of multiple H$_2$ lines to those predicted by models of shocked molecular hydrogen gas  and conclude that the emission is consistent with shock heating.  We use these data to estimate the mass and physical conditions of the emitting gas residing within the cleared region of the GG-Tau~A's circumbinary disk.  Finally, we discuss the phenomenon of shocked infalling gas in a binary system and the implications for similar systems such as transitional disks that potentially harbor planetary companions in their inner dust-cleared gaps.

\section{Observations}


Observations were obtained using the Near IR Integral Field Spectrograph (NIFS) at the Gemini North Frederick C. Gillette Telescope on Mauna Kea, Hawaii.   NIFS is an image slicing IFU fed by Gemini's Near IR adaptive optics (AO) system, Altair, that is used to obtain 3-dimensional x, y, wavelength ($\lambda$) spectral cubes at spatial resolutions of $\le$0\ptsec1 with a 2 pixel spectral resolving power of R$\sim$5300 at 2.2~$\mu$m \citep{mcgr2003}.  The NIFS spatial field is 3$''\times3''$ in size and the individual IFU pixels provide 0\ptsec1$\times$0.$''$04 spatial sampling.  Data were obtained at the standard $K$-band wavelength setting for a spectral range of 2.010-2.455~$\mu$m.  GG~Tau~A was bright enough to serve as its own wavefront reference stars for the facility adaptive optics system to guide at a 1000~Hz rate.  All observations used for the final co-added datacube were acquired in natural seeing of better than $\sim$0\ptsec85 for excellent AO correction. 

The data for this project was acquired in queue observations during Fall 2009 under program ID GN-2009B-Q-40 through thin cirrus and good seeing. Observations of GG~Tau~A were executed over four different nights.  The data reduction and analysis of the IFU datacubes on GG~Tau was identical to our past integral field spectroscopic studies \citep{beck2008,beck2010}.  The individual reduced GG~Tau~A datacubes were re-sampled to 0\ptsec04$\times$0\ptsec04 size pixels in the spatial dimensions.  The exposures each had 40 second integration times, and the data was acquired with small dithers that were the same size as the width of the NIFS IFU slicing mirrors.  The purpose of dithering the stars at integer slice mirror widths was to sample the stellar point spread function (PSF) in the cross dispersion direction in different pixels to improve the spatial sampling and mitigate the effect of the NIFS rectangular spatial elements.  The final datacube on GG~Tau~A was constructed by shifting-and-adding 78 individual dithered exposures for a total on-source exposure time of 3210 seconds.  To verify the $K$-band flux of the GG~Tau~A system and calibrate the imaging spectroscopy, supporting photometry was acquired in Sep. 2009 using the WIYN High-Resolution Infrared Camera \citep[WHIRC;][]{meix2010}, using the $K_s$ filter and the United Kingdom Infrared Telescope (UKIRT) faint standard FS~18 for flux calibration.   The derived $K_s$ magnitude for the combined GG~Tau~A brightness is 7.7$\pm$0.1, which is consistent with all published photometry of the system.

By inspecting the wavelength calibration and sky emission line positions, we estimate that the absolute velocity accuracy of our individual IFU cubes for a given night of observations is $\sim$9-12~km~s$^{-1}$ (roughly one third of a spectral pixel, or one sixth of a resolution element).  For our observations, the accuracy of the NIFS IFU grating wheel repositioning was repeatable only to $\sim\pm$0.75 pixels over the four nights of our observations.  Hence, because of very weak signal from H$_2$ emission in GG~Tau~A, data acquired over multiple nights was not subsampled and shifted and co-added in the wavelength dimension to preserve velocity resolution.  As such, we estimate that our final emission line velocity centroid accuracy is affected by the grating repositioning and the straight co-addition of data that is slightly shifted in wavelength.  As a result, the estimated absolute velocity accuracy of the GG~Tau~A data is at the level of 0.7 to 1.0 pixels, or 20-28~km~s$^{-1}$.  




\section{Results}

Figure~1 presents the image of 2.12~$\mu$m {\it v}~=~1-0~S(1) H$_2$ emission in the environment of GG~Tau~A.  This image was constructed by summing six spectral pixels centered on the H$_2$ emission at each spatial location in the field, after the continuum level was subtracted using a linear fit to the stellar continuum flux on each side of the emission feature.  GG~Tau~Aa, the southern star, is centered in the image.  Overplotted in blue are the 2$\mu$m continuum contours for both sources.  The H$_2$ emission is distributed around the binary, with obvious extensions of the emission to $\sim$70~AU (0\ptsec5) distances from GG~Tau~Aa.   The strongest H$_2$ emission arises from north-east of the system, about 0\ptsec2 away from the peak continuum position of GG~Tau~Ab.  Given the projected separation of the GG~Tau~A binary, the H$_2$ gas emission arises from dynamically unstable locations near the stars.   In the image, the apparent decrease in H$_2$ emission strength at the positions of the stars is likely not genuine, it results from decreased sensitivity to the faint H$_2$ over the very bright stellar flux.  If H$_2$ at the level of the peak emission seen in Figure 1 existed at the location of GG Tau Aa, it would be less than 0.5\% of the continuum flux and would be extremely difficult to detect after the continuum subtraction process.

A binned 2x2 image of the H$_2$ emission from the environment of GG~Tau~A is shown in Figure~2a.   In this image, the brightest H$_2$ is saturated to emphasize the low level emission in regions surrounding the young binary.  H$_2$ ``streamers" of emission are seen extending out to $\sim$0\ptsec9 distances from the central stars.  Three positions are highlighted in the image - Positions A, B and C. To the right in Figure~2b, the 1-D spatially integrated spectra (from 0\ptsec12 radius apertures) of the 2.12~$\mu$m {\it v}~=~1-0~S(1) line are presented.  The integrated signal-to-noise ratios are $\sim$24, 4 and 16 for the A, B and C spectra, respectively.  The velocity of the H$_2$ emission in the A and B spectra are centered at $\sim$-10~km~s$^{-1}$.  Given the $\sim\pm$20-28~km~s$^{-1}$ absolute accuracy of our velocity calibration, this is not a statistically significant difference from the systemic radial velocity of GG~Tau~A of 18~km~s$^{-1}$ \citep{hart1986}.  These results are consistent with the previous high spectral resolution (R$\sim$60,000) detection of H$_2$ emission from the GG~Tau~A system \citep{bary2003}.  The spectrum extracted at the C position has a double-peaked profile shape, with a significant redshifted contribution to the line flux.

Figure~3 presents a spectrum of the H$_2$ emission summed from spatial locations that have greater than 4.0$\times$10$^{-17}$ W~m$^{-2}$ line flux in the {\it v}~=~1-0~S(1) line.   In addition to the {\it v}~=~1-0~S(1) line at 2.12~$\mu$m, five other H$_2$ emission transitions are detected, including the {\it v}~=~2-1~S(1) line at 2.24~$\mu$m.   The measured value of the {\it v}~=~2-1~S(1) /{\it v}~=~1-0~S(1) line ratio is commonly used to determine the physical conditions of the emitting H$_2$ gas.  Assuming that the gas is in LTE, the Boltzmann relation and line ratio provide an estimate of the temperature of the emitting gas.  For the spectra in spatial locations nearby to GG~Tau~A (Figure 3), we measure a value of $\sim$0.05$\pm$0.01 for this line ratio and determine a corresponding gas temperature of $\sim$~1700~$\pm$100~K.  At this temperature and average $\sim$40~AU distance from the central stars, it is unlikely that this gas is heated by blackbody emission from the stars.  Given that the line ratio is also consistent with shock heating or high-energy photo-excitation in a dense medium ({\it n$_H$} $\ge$ 10$^4$~cm$^{-3}$), both processes are potential mechanisms for stimulating the observed emission.  However, given the spatial extent of the emission, the lack of detectable X-ray emission and relatively weak UV~flux from the binary, we conclude that shock heating is the most likely excitation mechanism for the H$_2$.

The detection of {\it v}~=~2-1~S(1) emission at $\sim$0\ptsec2 distances from the central stars is a very high-contrast measurement.  The peak flux in the brightest region of {\it v}~=~1-0~S(1) emission is 0.6\% of the continuum flux at the position of GG~Tau~Ab.  The detected {\it v}~=~2-1~S(1) emission is 5\% of the flux of the 1-0~S(1) line.  This discrete line emission is detected at 5$\sigma$ confidence, with a contrast ratio of  3$\times$10$^{-4}$ with respect to the continuum flux level of GG~Tau~Ab.    Our total emission line sensitivity is better than this because the stellar PSF of only the closer GG~Tau~Ab star in the binary is considered.  We estimate the overall sensitivity at the $\sim$0\ptsec2 distance from both stellar PSFs to be $\approx$8$\times10^{-5}$.   At this sensitivity level, we are successfully probing the parameter space of the next generation planet survey instruments, such as those using extreme coronographic adaptive optics and non-redundant masking techniques \citep{beic2010}.  This is very difficult, and is made possible primarily because we are searching for discrete, narrow emission lines where the adaptive optics halo structure can be accurately subtracted from the stellar point spread function over the small wavelength range.   Our detection of {\it v}~=~2-1~S(1) emission in the GG~Tau~A system highlights the success of using adaptive optics fed IFU spectroscopy to spatially resolve emission line morphologies within the inner 30~AU circumstellar environments of bright young stars.

\section{Discussion}

The GG~Tau system was first spatially resolved into a hierarchical quadruple by \citet{lein1991}.  At the time of those observations, the stars in the GG~Tau~A component of the quadruple had a spatial separation of 0\ptsec255 at a position angle of +9$^{\circ}$.  Our observations, acquired $\approx$20 years later, show a system separation of 0.251$\pm$0.008 at position angle 334$\pm$1 (measured east of north).  In the twenty years since its discovery as a binary, the GG~Tau~A stars have changed their orientation by nearly 35 degrees, just under one tenth of the total angular orbit \citep{kohl2011}.  The fact that the separation has remained largely constant over a wide arc of the orbit means that the system is viewed close to pole on and either the orbit is nearly circular, or the components are near periastron where the apparent separation would vary little \citep{kohl2011}.   The masses of the two stars in the GG~Tau~A system are estimated at $\sim$0.8 and $\sim$0.7 $M_{\odot}$ \citep[total system mass of 1.28$\pm$0.08;][]{whit1999,dutr1994,guil1999}.  \citet{hart1986} found the combined GG~Tau~A system to have a vsini of less than 10~km~s$^{-1}$, implying that the stellar components may also be viewed close to pole-on.   Though, in order to explain the large $\sim$180~AU extent of the inner edge of the circumbinary ring, dynamical modeling of the system implies that the orbital eccentricity of GG~Tau~A may lie in the range of 0.2-0.4 and the stellar binary orbit may not be coplanar with the circumbinary ring \citep{beus2005,kohl2011}.   Yet, when the full GG~Tau system is modeled as a quadruple then the need for non-coplanarity of GG~Tau~A with the circumbinary ring is removed \citep{beus2006}.

The peak of the detected molecular hydrogen emission in the GG~Tau~A system is located at a position that is $\sim$0\ptsec2-0\ptsec25 away from both GG~Tau~Aa and Ab.  Given the separation of the components of the binary of 0\ptsec25, the tidally truncated inner circumstellar disks will not extend to this distance from the stellar positions.  Hence, H$_2$ detected at this position is too extended to be associated with either of the circumstellar disks that exist around the stars.    Hydrodynamic models simulating the clearing of the inner regions around young star binaries clearly suggest that material at the observed location of H$_2$ emission should be dynamically cleared on timescales of a few hundred years \citep{arty1996,gunt2002}.  Though, the presence of material within the dynamically unstable region is not unexpected, given that GG Tau system possesses several indicators of accretion activity and the inner disks must be replenished to persist.

The circumbinary ring encircling GG~Tau~A was first detected in scattered light by \citet{rodd1996}.  Since the time of the first scattered light images, higher sensitivity observations have revealed an inner distribution of dust within the circumbinary ring of GG~Tau~A, extending to $\sim$0$\ptsec$5 north of the inner binary \citep{duch2004}.  This inner dust distribution corresponds in spatial location to the northern extension of H$_2$ emission that we see.  Our detected extended H$_2$ emission can not merely be scattered line flux from a compact, centrally excited H$_2$ emission environment because a central emission region would need to be hundreds to thousands of times stronger than our detected extended line emission region.   To produce our measured extended H$_2$ line flux, a central emission region would need to be at a level of more than a 30\% of the continuum, and would be easily detectable over the central stellar flux.   Moreover, if our detected H$_2$ line emission arose from scattered line emission, we would also expect to detect H$_2$ at the circumbinary ring, which is the location of the strongest and most easily detected scattered flux.  We do not detect appreciable H$_2$ emission from the circumbinary ring, to a detection limit of 8.0$\times 10^{-19}$ W/m$^2$/pixel.  The brightest region of H$_2$ emission we detect seems to be spatially coincident with the inner region of dust-scattered light revealed in broad-band imaging, though the H$_2$ emission must be excited locally within this environment rather than scattered. 

Various high contrast observations of the GG Tau A system have reported the detection of ``spokes" or ``streamers" of material extending from the inner edge of the circumbinary ring inward toward the central binary system.  These observations have been met with some level of skepticism because while ground-based studies seem to show some weak evidence for the streamers, images from the {\it Hubble Space Telescope} typically did not \citep{kris2005, mcca2002}.   For the first time, \citet{piet2011} detected a possible streamer of dust emission in the mm wavelength map of the circumbinary ring around GG~Tau~A.  This streamer of dust was detected at a $\sim$2$\sigma$ level of confidence, and it seems to extend from the northern edge of the circumbinary ring inward to the north side of the central binary.   Figure~4 presents an image of our H$_2$ gas emission around the GG~Tau~A system with the 1.1~mm dust continuum contours from \citet{piet2011} overplotted.    Our H$_2$ map also shows evidence of "streamers" of emission, and several of the H$_2$ extensions are detected at high signal-to-noise.  Interestingly, the location of the inner edge of the mm dust streamer corresponds precisely to the location where we detect the peak emission in molecular hydrogen gas.   The presence of the peak H$_2$ emission at a coincident location with the inner dust streamer lends further support that the streamer in the dust map is a real feature.  Additionally, \citet{guil2000} report the existence of CO emission from these regions, which would also support the existence of streaming gas interior to the circumbinary dust ring.

In the environments of young stars, most excitation mechanisms considered for ro-vibrational H$_2$ excitation involve stimulation by X-ray or UV~flux from the central star, or shock excitation by material in an outflow.  \citet{bary2003} using the X-ray excitation models of \citet{malo1996}, ignored the complexities of the binary system and assumed a centrally located X-ray/UV source incident on material located in a standard disk model \citep{glas2000} to predict a range of plausible H$_2$ line fluxes for the GG~Tau~A system.  Using log~L$_{x}$~=~29.4~ergs~s$^{-1}$\footnote{\citet{bary2003} did not note that this value was taken as an upper limit since GG~Tau~A was not detected by ROSAT}, \citet{bary2003}assumed that the X-ray energy was deposited in the upper atmosphere of a singular disk between 10 and 30 AU from the X-ray source.  In the best case scenario, the X-rays were capable of stimulating roughly 70\% of the observed H$_2$ line flux in a high-resolution spectrum of GG~Tau~A.  \citet{bary2003} further assumed the UV flux from GG~Tau to be similar to that of TW~Hya and estimated the contribution of UV fluorescent H$_2$ emission to be substantially less than the X-rays.  However, they concluded that the combination of the limit of X-ray ionization and UV fluorescence might account for a significant portion of the H$_2$ line flux in the GG~Tau~A system.  With the spatial resolution and wavelength coverage of NIFS, we find that the gas is not distributed in the manner predicted by high-energy photon disk excitation models described by \citet{bary2003,bary2008}.  The H$_2$ emitting gas extends farther from the sources than predicted and appears to display a morphology better described by gas participating in an accretion streamer than the central compact circumstellar disks illuminated by X-rays.  In addition, the {\it v}~=~2-1~S(1) / {\it v}~=1-0 S(1) line ratio of 0.05$\pm$0.01, measured at $\sim$40~AU distances corresponds to $T$~$\sim$1700$\pm$100~K.   This ratio can be accurately modeled by C-type shocks with reasonable pre-shock gas densities and shock velocities \citep[see below;][]{lebo2002}.  In light of these considerations, the lack of detectable X-ray flux from the GG~Tau~A system, and the spatial coincidence of the northern H$_2$ gas with the possible dust stream (Figure~4), we conclude that this gas is most likely being shock excited by infalling material crashing into gas distributed throughout the circumstellar domain than stimulated by high-energy photons. 

The measured emission line flux of the {\it v}~=~1-0~S(1) transition and the estimated gas excitation temperature coupled with the partition function analysis from \citet{herb1996} allow us to determine the molecular hydrogen column density:

\begin{equation}
N_{tot}=4.55\times10^{16} \cdot T \cdot  (e^{6000/T}-1)e^{956/T} \cdot I_{\lambda} \cdot 10^{0.4\rm{A}_{\lambda}}
\end{equation}

\noindent where {\it I}$_{\lambda}$ is the measured line intensity (in erg~cm$^{-2}$~s$^{-1}$~sr$^{-1}$), $T$ is the gas temperature (assuming LTE), and A$_{\lambda}$ is the extinction to the emitting region, which we adopt to be 0.7 \citep{whit1999}.  The derived column density of H$_2$ in the brightest knot of emission is N$_{tot}$~=~$1.6\times10^{15} $~cm$^{-2}$.  The H$_2$ emitting region sampled here was a square pixel with a 0\ptsec1 size.  Hence these dimensions and the calculated $N_{tot}$ implies a mass of  $\sim10^{-12}$~M$_{\odot}$ in molecular hydrogen in this aperture, assuming a 140~pc distance to the Taurus molecular cloud.   Assuming comparable column densities in regions of the H$_2$ circumbinary material encompassing the system out to $\sim$0\ptsec3 distances, we estimate that the mass in H$_2$ around GG~Tau~A is at least several $\times10^{-11}$~M$_{\odot}$.   Using the stellar parameters from GG~Tau~Aa and Ab from \citet{whit1999} with the accretion rate relations from \citet{muze1998}, we have used our measured Br$\gamma$ line luminosity to derive mass accretion rates onto the stars.   These accretion rates are 2.7$\times$10$^{-8}$~M$_{\odot}$~yr$^{-1}$ and 2.4$\times$10$^{-8}$~M$_{\odot}$~yr$^{-1}$ for GG~Tau~Aa and Ab respectively, consistent with past mass accretion rate estimates for the system \citep{whit1999}.  Not only does dynamical theory predict that the H$_2$ should not be stable at our detected locations, but the mass accretion rate onto the stars is several orders of magnitude larger than the estimated mass of emitting H$_2$ in the circumbinary environment.  At this level of mass accretion, the extended H$_2$ material must be continually replenished or it would be accreted rapidly onto the central disks and stars.

We have compared the detected fluxes of the {\it v}~=~1-0~S(0), S(1), S(2) and {\it v}~=~2-1~S(1) H$_2$ lines from the spectra of GG~Tau (Figure~3) with the C-type shock models of \citet{lebo2002}.  We probed a range of pre-shock medium densities from $10^{3}$ to $10^{7}$~cm$^{-3}$ and shock velocities from 10~km~s$^{-1}$ to 70~km~s$^{-1}$.  The very slow shocks ($<$10~km~$^{-1}$) are ruled out with confidence, at these low velocities the shocks cannot produce the measured excitation temperatures greater than 1000~K.  Extremely fast high-density shocks also cannot reproduce the measured line fluxes, as these models result in greater excitation temperatures measured by the {\it v}~=~2-1~S(1) / {\it v}~=~1-0~S(1) line ratio.  We find that the optimal solutions in the models are for a pre-shock medium in the density range of $10^3 - 10^5$~cm$^{-3}$ with a shock velocity of 20-50~km~s$^{-1}$.  In GG Tau A, the accretion infall shock velocity would result from a combination of the gas free-fall velocity onto the system ($\sim$10~km~s$^{-1}$) with a component from the Keplerian motion of material in the inner system. With these constraints, only the model calculations of shock velocities in the lower 20-30~km~s$^{-1}$ range would be consistent with the expected gas accretion kinematics for GG~Tau~A.   

We speculate that H$_2$ can also be stimulated into emission by mass accretion infall in other systems, such as young star binaries and systems with ``transitional disks" that have dust cleared disk gaps that are thought to trace the onset of planet formation.  In these latter systems,  gas and dust from the outer distribution of disk material could accrete onto the inner regions, and H$_2$ might be stimulated into emission from the shock excited free-falling material, which could have velocities in the 20-40~km~s$^{-1}$ range.  For example, the GM~Aur transition disk system has a dust-cleared gap in the inner disk that extends out to a distance of $\sim$20-25~AU from the central star \citep{dutr2008,hugh2009}.  Accretion streams of gas that free-fall from the $\sim$20~AU outer ring onto an optically thin inner disk in the $\le$1~AU region \citep{espa2010} could have a velocity at the 30-40~km~s$^{-1}$ level, which would be fast enough to shock-excite the H$_2$ into emission.

In addition to the peak of H$_2$ emission coincident with the inner dust streamer, extensions of H$_2$ emission exist out to $\sim$0\ptsec7 (100~AU) separations from the central stars in GG~Tau~A (Position~B in Figure~2).  The detected H$_2$ gas itself seems to exhibit ``streamers" of emission that radiate outward from the central system.  More sensitive data at both mm and near-infrared wavelengths could clarify the correlation between the streamers and the dust and gas species present within the system.  Further study of the spatially resolved material apparent in the GG~Tau~A ring system would lead to a more detailed understanding of the general physics of mass accretion onto a central stellar binary.  While this knowledge is directly applicable to other young star binaries, it also can serve to constrain the Keplerian motion in a broader base of models such as those that describe mass accretion onto proto-planets within circumstellar disks, as well as binary black holes in galaxy mergers and binary stellar remnants.
	


\acknowledgments

We are grateful to the Gemini North Observatory queue observers who acquired the data for this project, and to our NOAO and Gemini contact scientists Knut Olsen and Richard McDermid, who helped prepare this program for execution.  Data for this project was acquired under Gemini Observatory program ID GN-2009B-Q-40.   This study is based on observations obtained at the Gemini Observatory, which is operated by the Association of Universities for Research in Astronomy, Inc., under a cooperative agreement with the NSF on behalf of the Gemini partnership: the National Science Foundation (United States), the Science and Technology Facilities Council (United Kingdom), the National Research Council (Canada), CONICYT (Chile), the Australian Research Council (Australia), Minist\'{e}rio da Ci\^{e}ncia e Tecnologia (Brazil) and Ministerio de Ciencia, Tecnolog\'{i}a e Innovaci\'{o}n Productiva (Argentina).

\clearpage



\begin{figure}
\epsscale{1.1}
\plotone{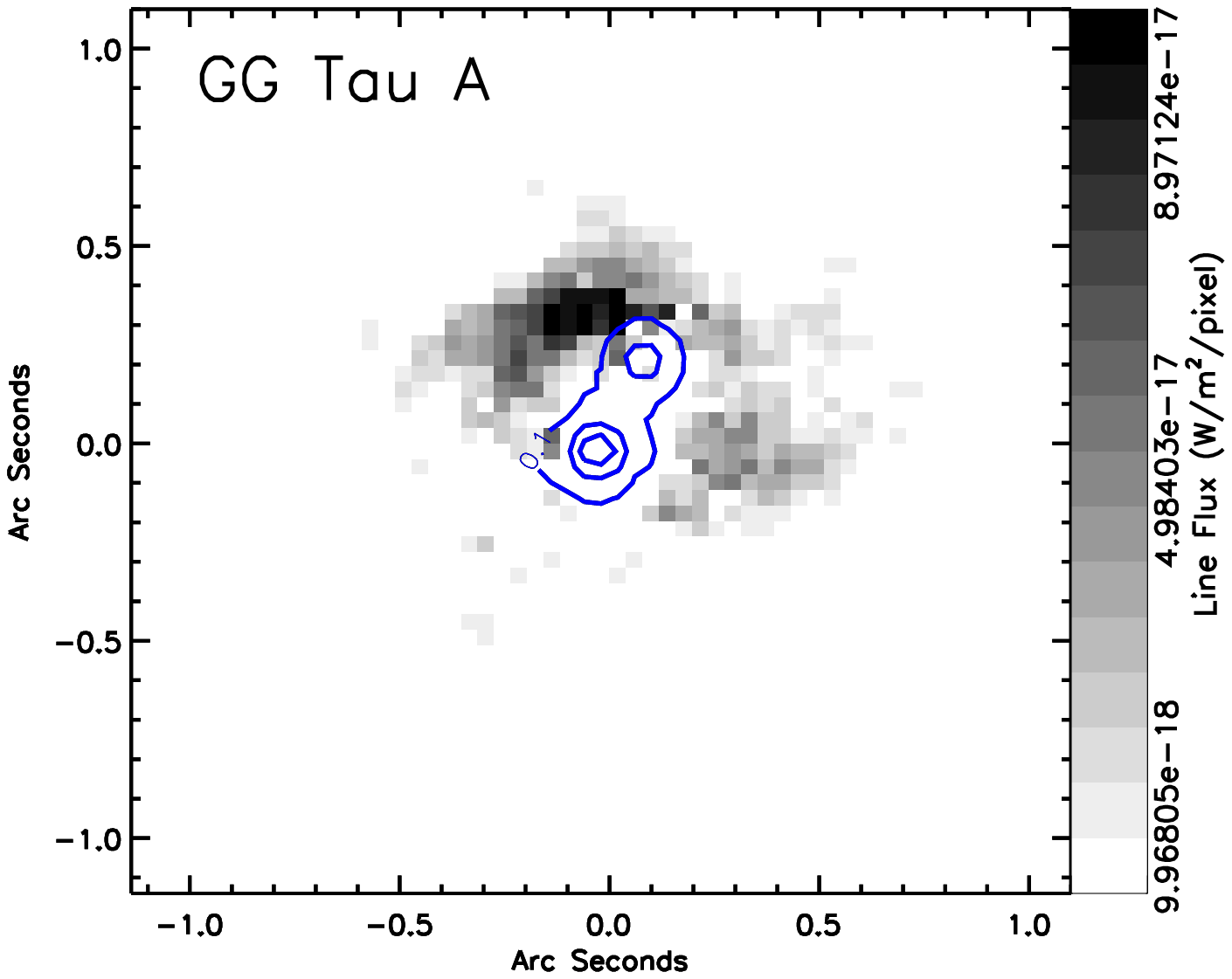}
\caption{A Continuum subtracted grey scale image of H$_2$ emission in the environment of GG Tau A binary, with continuum contours at 0.1, 0.4 and 0.7 times the peak continuum flux of GG Tau Aa overplotted in blue.  The 0.7 and 0.4 contour levels were selected to highlight the continuum flux positions of the GG Tau Aa and Ab stars in this young binary.  The H$_2$ emission is distributed around the binary, with extensions to the emission out to $\sim$100 AU (0.$"$7) distances.  The apparent decrease in H$_2$ emission at the positions of the stars is likely not genuine, it results from uncertainty in the continuum subtraction process to detect the faint H$_2$ over the bright stellar flux.  \label{fig1}}
\end{figure}

\clearpage


\begin{figure}
\epsscale{1.18}
\plotone{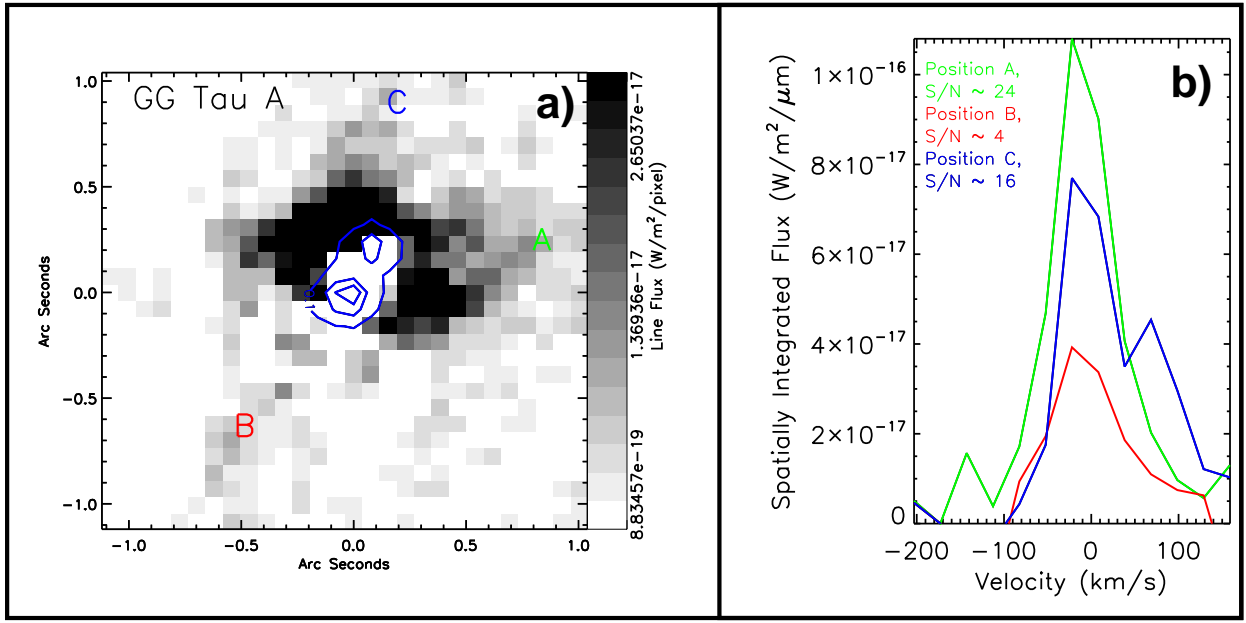}
\caption{To emphasize the detection of low level line emission, a binned 2x2 image of the H$_2$ emission from the environment of GG Tau A is presented in a).   In this image the display is saturated to show the emission surrounding the young star binary.   Three positions - Position A, B and C - are highlighted in color in the image. To the right in b), the 1-D spatially integrated spectra of the 2.12$\mu$m {\it v} = 1-0 S(1) line at the A, B and C positions are presented.  The H$_2$ line emission is detected at significant signal-to-noise at spatial positions of more than 100AU from the central stars.   \label{fig2}}
\end{figure}


\begin{figure}
\includegraphics[angle=90,scale=.75]{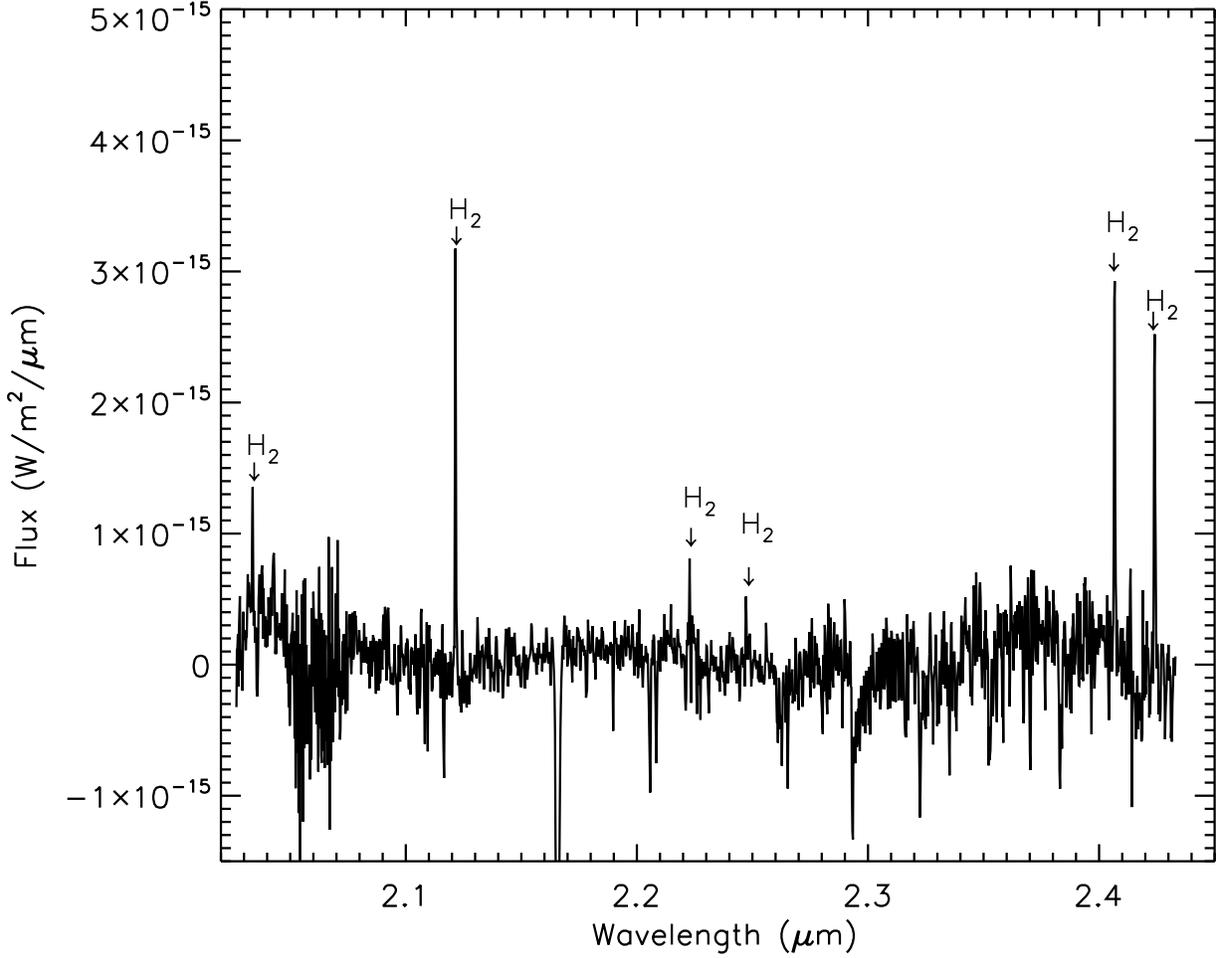}
\caption{A spectrum of the H$_2$ emission summed from regions of flux that are brighter than 4.0$\times$10$^{-17}$ W~m$^{-2}$ (as seen in Figure 1).   In addition to the {\it v} = 1-0 S(1) line at 2.12$\mu$m, five other H$_2$ emission transitions are detected, including the {\it v} = 2-1 S(1) line at 2.24$\mu$m.  The  {\it v} = 2-1 S(1)/{\it v} = 1-0 S(1) ratio we detect from the gas around GG Tau A is $\sim$0.05$\pm$0.01, implying the presence of warm, T$_{ex}\sim$1700K gas at extended $\sim$40 AU distances from the system.  Residual stellar photospheric absorption features of Al+Mg (2.11$\mu$m), Na (2.206/2.208$\mu$m), Ca (2.26 $\mu$m) and CO (2.29 - 2.38$\mu$m) are seen in this spectrum.}
\end{figure}

\begin{figure}
\includegraphics[scale=1.1]{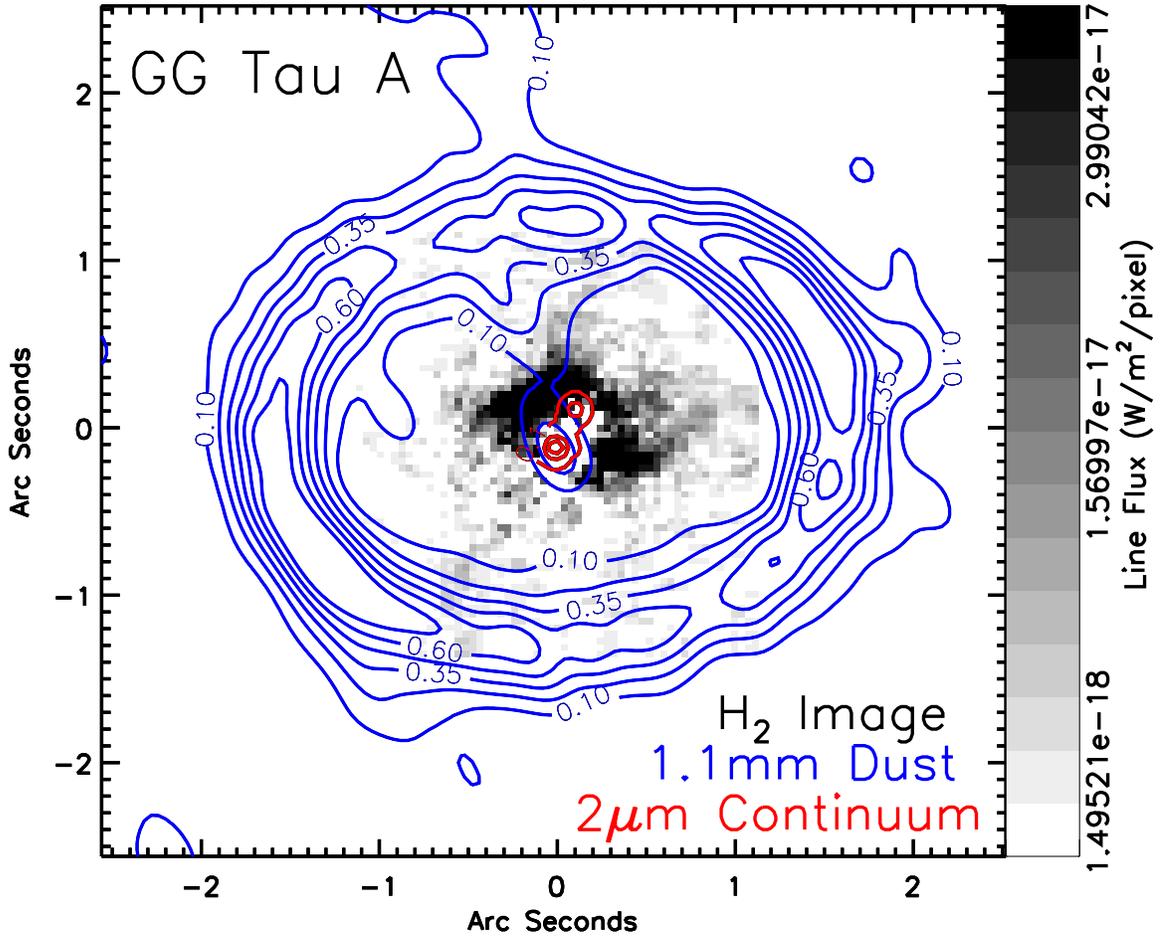}
\caption{  The continuum subtracted H$_2$ image in the environment of GG Tau A (grey) with overplotted contours of the 1.1mm dust continuum emission (blue) and the 2$\mu$m continuum flux (red).  The brightest peak of molecular hydrogen emission corresponds to the spatial location of the dust "streamer" that is seen to extend within the circumbinary ring, which implies that both features are real and the H$_2$ may be shock excited by mass accretion infall onto the binary stars.}
\end{figure}

\end{document}